\documentclass{article}

\newcommand*{\Mname}{{FlashCP}}

\newcommand{\BlueComment}[1]{\textcolor{blue}{\footnotesize{// #1}}}

\usepackage{microtype}
\usepackage{graphicx}
\usepackage{subcaption}
\usepackage{booktabs} 

\usepackage{hyperref}

\usepackage[preprint]{icml2026}

\usepackage{amsmath}
\usepackage{amssymb}
\usepackage{mathtools}
\usepackage{amsthm}

\usepackage[capitalize,noabbrev]{cleveref}

\theoremstyle{plain}

\theoremstyle{definition}

\theoremstyle{remark}

\usepackage[textsize=tiny]{todonotes}

\icmltitlerunning{FlashCP: Load-Balanced Communication-Efficient Context Parallelism for LLM Training}

\begin{document}

\twocolumn[
  \icmltitle{FlashCP: Load-Balanced Communication-Efficient \\ Context Parallelism for LLM Training}



  \icmlsetsymbol{equal}{*}

  \begin{icmlauthorlist}
    \icmlauthor{Zheng Wang}{ucsd}
    \icmlauthor{Eric Liu}{usc}
    \icmlauthor{Linan Jiang}{ucsd}
    \icmlauthor{Zhongkai Yu}{ucsd}
    \icmlauthor{Zaifeng Pan}{ucsd}
    \icmlauthor{Yue Guan}{ucsd}
    \icmlauthor{Yuke Wang}{rice}
    \icmlauthor{Yufei Ding}{ucsd}
  \end{icmlauthorlist}

  \icmlaffiliation{ucsd}{University of California San Diego, La Jolla, USA}
  \icmlaffiliation{usc}{University of Southern California, Los Angeles, USA}
  \icmlaffiliation{rice}{Rice University, Houston, USA}

  \icmlcorrespondingauthor{Zheng Wang}{zhw100@ucsd.edu}

  \icmlkeywords{Machine Learning, ICML}

  \vskip 0.3in
]



\printAffiliationsAndNotice{}  

\begin{abstract}
Context parallelism (CP) is essential for training large-scale, long-context language models, as it partitions sequences to reduce memory overhead. However, existing CP methods suffer from workload imbalance, inefficient kernels, and redundant communication due to static sequence sharding and key-value (KV) tensor communication. We present \textit{FlashCP}, a load-balanced and communication-efficient framework for CP training. \textit{FlashCP} introduces a sharding-aware communication mechanism to eliminate redundant KV communication and proposes a novel Whole-Doc sharding strategy that maximizes communication savings while maintaining balanced workloads. To efficiently combine Whole-Doc and Per-Doc sharding, \textit{FlashCP} further designs a heuristic algorithm to search for near-optimal sharding plans. Extensive experiments show that \textit{FlashCP} achieves up to $1.63\times$ speedup over state-of-the-art CP frameworks across diverse datasets.
\end{abstract}

\section{Introduction}
Large language model (LLM) has demonstrate impressive capability in many tasks like translation~\cite{zhang2023prompting}, reasoning~\cite{guo2025deepseek}, and coding~\cite{wei2023copiloting, nijkamp2022codegen}. 
The remarkable potential of LLMs has driven a growing trend among leading technology companies to developing increasingly larger-scale models with extended context windows, continually pushing the boundaries of LLM capabilities~\cite{achiam2023gpt, grok3beta, dubey2024llama, team2023gemini}.
For example, the context window size of the Llama model series evolves from 4K tokens in Llama2~\cite{touvron2023llama}, to 128K tokens in Llama3~\cite{dubey2024llama}, and finally reaching 10 million tokens in Llama4~\cite{singh2025meta}.

This increase in context window size introduces significant challenges for LLM training, primarily due to the growth in the activation size, which scales proportionally with the sequence length~\cite{korthikanti2023reducing}.
To address this challenge, in addition to the traditional 3D parallelism techniques (data parallelism, pipeline parallelism, and tensor parallelism)~\cite{shoeybi2019megatron, narayanan2021efficient, team2020deepspeed}, a new dimension called Context Parallelism (CP) has been introduced~\cite{nvidia2023megatron}. 
With CP, the inputs and activations are partitioned along the sequence length dimension and distributed across CP workers, allowing attention computation to run in parallel on different devices. This approach effectively reduces the memory consumption of attention layer activations on each device, offering a promising solution to train LLMs with large context windows.

Although context parallelism is essential for efficient long-context LLM training, achieving optimal performance remains challenging, and all existing solutions fall short in some aspects.
\textbf{First}, balancing the attention workload across CP workers is difficult because the per-token attention computation varies with sequence position, leading to imbalanced workload distribution across GPUs.
\textbf{Second}, CP introduces significant communication overhead due to the partitioning and distribution of the input sequence. Each token’s attention depends on all preceding tokens, requiring the communication of KV tensors across CP workers.
\textbf{Third}, maintaining high computation efficiency is nontrivial. Efficient attention kernels such as FlashAttention rely on sufficiently large query and KV lengths to achieve high GPU utilization, which is harder to sustain when the input is partitioned.
Addressing any one of these challenges in isolation is relatively straightforward. However, simultaneously resolving all of them to achieve optimal performance is difficult due to the inherent trade-offs between these aspects. 
For example, a more fine-grained partitioning can improve workload balance but shortens each sequence shard, decreasing kernel efficiency~\cite{wang2025wlb}. Similarly, communication overhead can be mitigated by overlapping computation and communication, but this approach requires splitting the attention kernel into several smaller kernels to compute partial results and introduces additional result-processing overhead~\cite{liu2023ring}.
Table~\ref{tab:cp_comparison} summarizes a comparison of mainstream CP approaches, all existing methods exhibit limitations in at least one dimension.

To address these challenges, we introduce \textit{\Mname}, a load-balanced and communication-efficient context parallelism framework for large-scale, long-context LLM training.
\textit{\Mname} holistically optimizes the sharding strategy and the communication flow in CP, effectively balancing workload distribution, maximizing attention kernel efficiency, and minimizing communication overhead to achieve near-optimal CP training performance.
Our key insight is that, rather than uniformly sharding and distributing all input documents, it is better to distribute the whole input document without sharding. Keeping the input document as a whole maximizes kernel efficiency and eliminates the need to communicate the KV tensor across GPUs.
To achieve balanced attention workload distribution with minimal document sharding, \textit{\Mname} incorporates several key optimizations:
\textbf{First}, to minimize the communication overhead, \textit{\Mname} employs a sharding-aware communication mechanism, communicating only the necessary portions of KV tensors required by each CP worker, thereby eliminating redundant data transfers.
\textbf{Second}, \textit{\Mname} proposes a novel \textit{Whole-Doc} sharding strategy, which keeps short documents as a whole on a single CP worker to reduce the required communication amount and adaptively shards the remaining documents to achieve balanced workload distribution and high attention kernel efficiency.
\textbf{Third}, \textit{\Mname} introduces a heuristic sharding algorithm that overcomes the NP-hardness of the sharding problem and efficiently searches for the near-optimal sharding plan that combines both \textit{Whole-Doc} and \textit{Per-Doc} sharding strategies.

\begin{table}[t]
\centering
\caption{Comparison of CP approaches.}
\resizebox{0.95\columnwidth}{!}{
\begin{tabular}{lccc}
\toprule
\textbf{Method} & \textbf{Balance} & \textbf{Communication} & \textbf{Kernel Eff.} \\
\midrule
Llama3 CP (Per-Seq) & Imbalanced & High & High \\
Per-Doc CP & Balanced & High & Low \\
Ring-Attn (Zigzag) & Balanced & Moderate & Low \\
\textit{\textbf{FlashCP}} & \textbf{Balanced} & \textbf{Low} & \textbf{High} \\
\bottomrule
\end{tabular}
}
\label{tab:cp_comparison}
\vspace{-10pt}
\end{table}

In summary, this paper makes the following contributions: 
\begin{itemize}
\item We reveal the limitations of existing CP frameworks, which fail to maintain high kernel efficiency and suffer from redundant KV tensor communication.
\item We propose \textit{\Mname}, which holistically optimizes input sharding and communication flow to achieve balanced workloads, reduced communication, and high kernel efficiency.
\item We compare \textit{\Mname} with state-of-the-art CP framework and observe up to $1.63\times$ speedup across various datasets.
\end{itemize}

\begin{figure}[t] 
    \centering
    \includegraphics[width=0.92\linewidth]{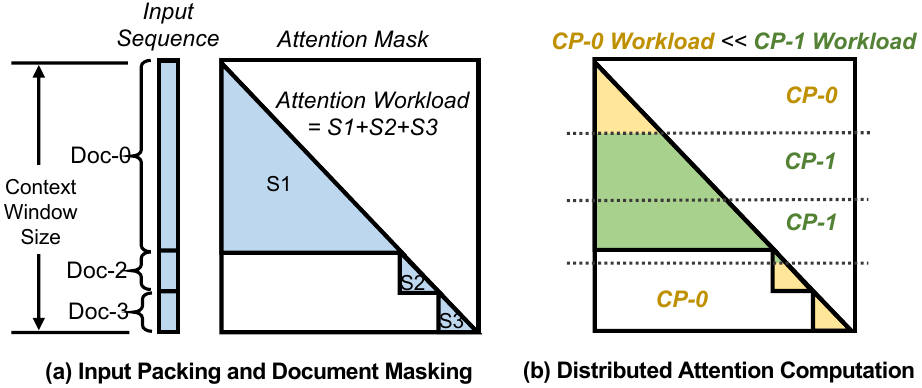}
    \caption{Illustration of input packing and distributed attention computation in context parallelism.}
    \label{fig:background}
    \vspace{-10pt}
\end{figure}

\section{Background and Motivation}

\subsection{Input Packing and Document Mask} 
Input documents in LLM training exhibit highly skewed length distributions~\cite{an2024does, jiang2024dynapipe, wang2025wlb}. Early approaches rely on zero-padding to align sequences within a batch~\cite{shoeybi2019megatron}. However, this padding approach introduces redundant computation, communication, and memory overhead.
To address this, input packing was proposed to concatenate multiple short documents into a single long sequence~\cite{zhao2024analysing, raffel2020exploring, krell2021efficient, wang2024packing}. Built upon input packing, the \textit{Document Mask} (also known as intra-document causal masking) was proposed to mask out cross-document attention computation and ensure correct attention behavior~\cite{flexattention, nvidianemo, zhao2024analysing, kundu2024enhancing}.
Figure~\ref{fig:background}(a) shows an example of input packing and document masking.
The combination of input packing and document masking has emerged as a widely adopted paradigm for large-scale, long-context LLM training~\cite{wang2025wlb, ge2025bytescale} and has been successfully applied in industry-scale models such as Llama3~\cite{dubey2024llama}.

\subsection{Context Parallelism} 
Context parallelism (CP) mitigates the large activation memory induced by long context windows by partitioning input sequences along the sequence-length dimension across multiple workers~\cite{nvidia2023megatron, gu2024loongtrain, dubey2024llama, wang2025wlb}. 
Each worker processes a subset of tokens and computes attention locally, while exchanging KV tensors to obtain the full attention context.
Figure~\ref{fig:background}(b) shows an example following the Llama3 CP implementation~\cite{dubey2024llama}, where the input sequence is split into $2 \times \text{CP\_size}$ shards, and each worker processes two shards.
This static sharding strategy can lead to significant workload imbalance across workers, highlighting the need for optimized document sharding and token distribution.

\begin{figure}
    \centering
    \includegraphics[width=0.9\linewidth]{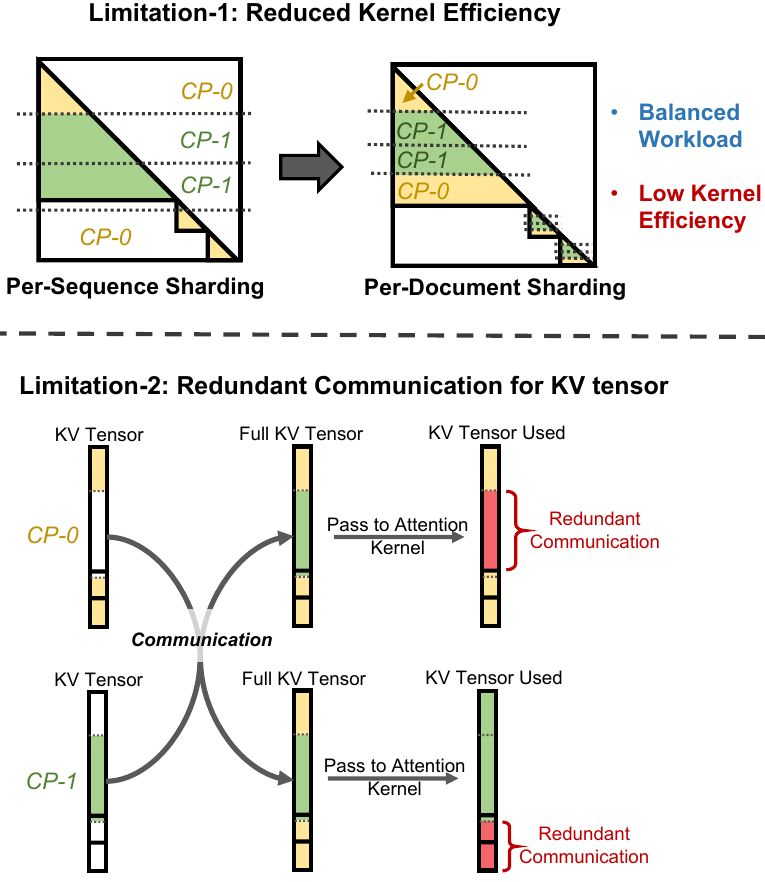}
    \caption{Limitations of existing CP training frameworks: (1) Existing CP implementations employ fine-grained per-document sharding to achieve balanced workload distribution, which reduces attention kernel efficiency; (2) Existing works communicate the full KV tensors across all CP workers, leading to redundant KV tensor transfers and unnecessary communication overhead.}
    \label{fig:limitation}
    \vspace{-10pt}
\end{figure}

\subsection{Limitation of Existing Works}
\begin{figure}
    \centering
    \includegraphics[width=0.95\linewidth]{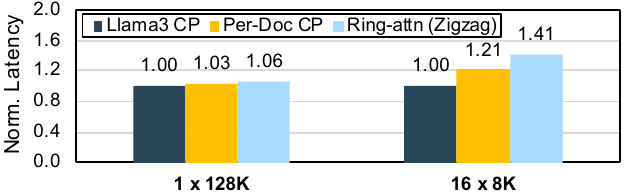}
    \caption{Kernel efficiency comparison: Fine-grained per-document sharding decreases attention kernel efficiency.}\label{fig:kernel_efficiency}
    \vspace{-10pt}
\end{figure}

The Llama3 CP~\cite{dubey2024llama} adopts a coarse-grained input sharding strategy that uniformly splits the entire input sequence. This approach often leads to severe workload imbalance across CP workers. To address this issue, two more advanced frameworks have been proposed: \textit{Per-Doc CP}~\cite{wang2025wlb} and \textit{Ring-Attn (Zigzag)}~\cite{ring2025zigzag}. 
Both frameworks employ a fine-grained, per-document sharding strategy, where each input document is divided into $2 \times \text{CP\_size}$ chunks. Chunks $i$ and $(2N - 1 - i)$ are then assigned to the $i$-th CP worker. As shown in Figure~\ref{fig:limitation}, this fine-grained document sharding strategy could achieve balanced workload distribution across CP workers.
Although both methods effectively achieve workload balance, they share several common limitations:

\textbf{Reduced Kernel Efficiency:} Fine-grained sharding causes each attention kernel to operate on shorter document shards, which decreases kernel efficiency~\cite{wang2025wlb}. To demonstrate this effect, we profiled kernel execution latency using two input patterns: one with a single document of length $128\text{K}$, and another with $16 \times 8\text{K}$ documents. As shown in Figure~\ref{fig:kernel_efficiency}, \textit{Per-Doc} sharding incurs noticeably higher attention latency, particularly for workloads dominated by short documents. \textit{Ring-Attn (Zigzag)} shows even higher latency since it computes attention block-by-block, introducing additional partial result aggregation overhead.

\textbf{Redundant Communication:} Although the two frameworks adopt different communication approaches, they both suffer from redundant communication. \textit{Per-Doc CP} relies on collective communication (e.g., \texttt{AllGather} and \texttt{ReduceScatter}), while \textit{Ring-Attn (Zigzag)} uses peer-to-peer (P2P) communication. In both cases, the entire KV tensors are transferred across all CP workers. However, each worker only requires a subset of the KV tensors to compute attention, resulting in unnecessary data transfer, as illustrated in Figure~\ref{fig:limitation}.

\section{{\Mname} Design}

\subsection{Optimization Goal and Problem Formulation}\label{sec:formulation}
Our goal is to determine an input sharding and distribution strategy that balances computation across CP workers while minimizing KV communication overhead and kernel efficiency degradation.
We consider context parallelism with CP size $N$ and context window $C$. Given an input sequence comprising $n$ documents $\mathbf{D} = [d_1, d_2, \cdots, d_n]$, where $d_i$ denotes the length of the $i$-th document, the documents are further partitioned into $m$ document shards $\mathbf{S} = [s_1, s_2, \cdots, s_m]$, where $s_i$ is the shard length. Each shard is also associated with a prefix length $p_i$, representing the number of tokens preceding its starting position.

\textbf{Input Shard Distribution.} For an input document shard $s_{i}$, its distribution is represented by an array of binary variables: $\mathbf{x}_i = [x_{i1}, x_{i2}, \cdots, x_{iN}]$, where $x_{ij} = 1$ if the shard $s_{i}$ is assigned to the $j$-th CP worker. Each shard is assigned to exactly one worker, enforced by the constraint:
\begin{equation}\label{eq:1}
\sum_{j=1}^{N} x_{ij} = 1,\ \ \forall i \in \{1, 2, \cdots, m\}
\end{equation}

\textbf{Equal Token Constraint.} To balance computation across CP workers, each worker must process the same number of tokens, since the cost of non-attention layers (e.g., FFN) scales linearly with token count. This requirement is enforced by the constraint:

\begin{equation}\label{eq:2}
\sum_{i=1}^{m}(x_{ij} \cdot s_{i})  = \frac{C}{N},\ \ \forall j \in \{1, 2, \cdots, N\}
\end{equation}

\begin{figure*}[h] 
    \centering
    \includegraphics[width=0.9\linewidth]{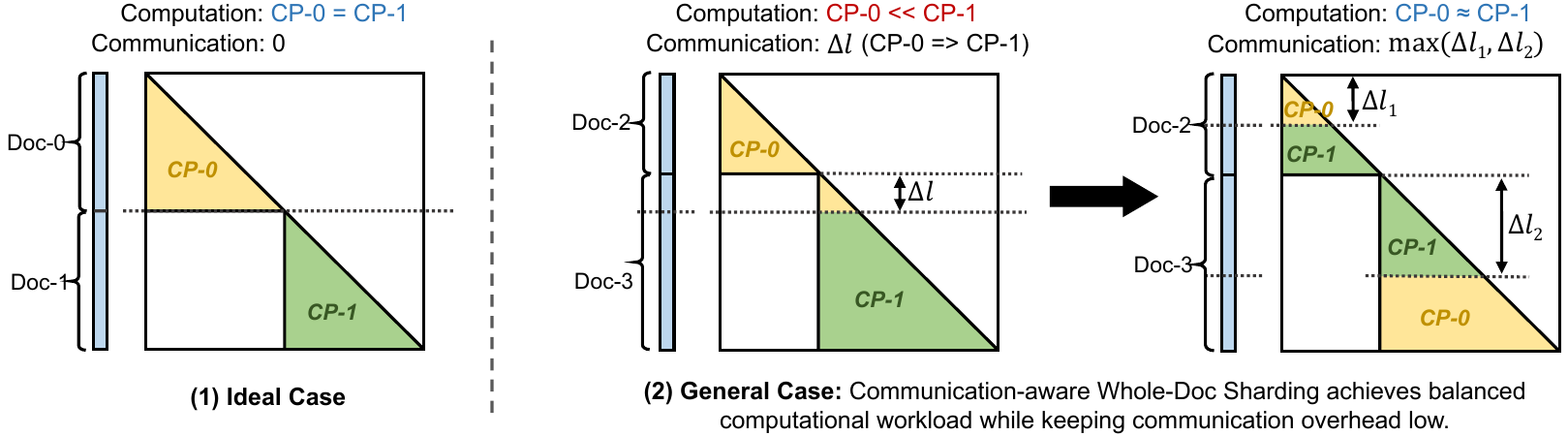}
    \caption{Illustration of communication-aware \textit{Whole-Doc} sharding.}
    \label{fig:whole_doc}
    \vspace{-10pt}
\end{figure*}

\textbf{Balancing Computation Workload.} Since training operates in a synchronized manner, the overall performance is bottlenecked by the slowest CP worker. Therefore, the objective is to minimize the maximum attention computation workload across all workers, subject to the constraints defined in Eq.~\ref{eq:1} and Eq.~\ref{eq:2}:

\begin{equation}\label{eq:3}
    \operatorname*{Minimize}_{\mathbf{X},\,\mathbf{S}}\left ( \max_{j \in {1, 2, \cdots, N}}  \sum_{i=1}^{m}(x_{ij} \cdot W_{i}) \right )
\end{equation}

Here, $W_{i}$ is the attention computational workload of the $i$-th input shard, which is computed as $W_{i}={(2\cdot p_{i} + s_i + 1)\times s_i}/{2}$.

\subsection{Dynamic Sharding-aware Communication} 
A missing component of the problem formulation in Section~\ref{sec:formulation} is the communication overhead. Existing CP implementations employ a \textbf{static communication pattern} that transfers the full KV tensors across all CP workers. The total communication volume along the critical path can be expressed as:

\begin{equation}
    4 \times \frac{\sum_{i=1}^{n} d_i}{N} \times H \times D \times (N - 1)
\end{equation}

Here, $H$ denotes the number of attention heads, $D$ the head dimension, and $N$ the number of CP workers. The factor of $4\times$ arises because communication is required for both the $K$ and $V$ tensors in both the forward and backward passes.
Since CP is typically applied across nodes, where communication bandwidth is significantly lower than that of intra-node NVLink, the communication overhead could become a significant bottleneck.

This approach is inefficient as it introduces redundant communication. To mitigate communication overhead in CP training, we propose a \textbf{dynamic sharding-aware communication} strategy, which adaptively determines the size of the communication buffer based on the sharding plan, thereby avoiding redundant KV tensor transfers.
Specifically, if an input document is assigned to a single CP worker, it is skipped for communication, since the CP worker already holds the full KV tensor for this document and can perform the attention computation locally.
For input documents that are sharded and distributed to multiple CP workers, the communication size is proportional to the maximum prefix length among all shards. This is because the Q tensor of each token only computes attention with the prefix part of the KV tensor in an input document.
Moreover, since each document shard may have a different prefix length, this results in zero padding in the communication buffer for each input document. To avoid this, rather than handling each document individually, we use a single continuous communication buffer and compact the KV tensors corresponding to the prefix portions of all document shards into the buffer. With our dynamic sharding-aware strategy, the communication size is reduced to:

\begin{equation}\label{eq:5}
    4 \times \left( \max_{j \in \{1, 2, \cdots, N\}}  \sum_{i \in \mathbf{\hat{S}}} (x_{ij} \cdot s_i) \right) \times H \times D \times (N-1)
\end{equation}

Here, $\mathbf{\hat{S}}$ denotes the set of input document shards excluding the last shard of each input document. An entire document that is not further divided is also treated as the last shard and, therefore, is not included in $\mathbf{\hat{S}}$. By applying the dynamic sharding-aware communication strategy, communication is reduced for two types of input shards: (1) documents that are fully assigned to a single CP worker, and (2) the last shard of each input document, achieving significant communication savings.

\subsection{Whole-Doc Sharding for Communication Savings} 
To maximize communication savings, we ideally want to each input document intact on a single CP worker, eliminating KV communication and preserving kernel efficiency by enabling single-kernel attention computation.
As shown in Figure~\ref{fig:whole_doc}(1), in the ideal case of two equal-length documents, the optimal strategy is to keep each document intact and assign it to a separate CP worker, achieving balanced computation and eliminating communication since each worker holds the full KV tensor. However, in practice, documents have varying lengths, making it difficult to preserve whole documents while evenly distributing workloads across CP workers.
In Figure~\ref{fig:whole_doc} (2), when the input contains documents of different lengths, the longer document must be split to satisfy the equal token constraint in Eq.~\ref{eq:2}. While this sharding plan incurs relatively low communication overhead (proportional to $\Delta l$), it introduces significant workload imbalance.
To address this challenge, we propose a \textbf{Communication-aware Whole-Doc Sharding} method, which adaptively shards documents to balance workload while minimizing communication overhead.
Specifically, instead of naively partitioning only the longest document to achieve equal token distribution across CP workers, we propose adaptively partitioning multiple documents. As shown in the right part of Figure~\ref{fig:whole_doc} (2), both \textit{Doc-2} and \textit{Doc-3} are partially partitioned to achieve balanced workloads and equal token counts per worker. Moreover, since only the lower segments of the documents need to be exchanged, the communication volume is bounded by $\max(\Delta l_1, \Delta l_2)$, significantly reducing overall communication overhead. This method is referred to as \textit{Whole-Doc Sharding}, as it aims to preserve each document as a whole and only applies adaptive sharding when necessary to balance the workload distribution.

\textbf{Combine \textit{Per-Doc} and \textit{Whole-Doc} Sharding.} Although \textit{Whole-Doc} sharding reduces communication overhead and can help achieve workload balance across CP workers, it cannot efficiently handle all combinations of input documents.
In certain cases, for example, when the input sequence contains an extremely long document that contributes the majority of tokens, \textit{Whole-Doc} sharding fails to achieve balanced workload distribution due to the significant disparity between the long document and the remaining short documents.
To address this, we design a hybrid sharding strategy that combines both \textit{Per-Doc} and \textit{Whole-Doc} sharding. Specifically, we select the appropriate sharding method for each document based on its length.
For extremely long documents that cause workload imbalance, \textit{Per-Doc} sharding is applied to evenly distribute computation, while maintaining high attention kernel efficiency due to sufficient token counts. In contrast, \textit{Whole-Doc} sharding is used for shorter documents to reduce communication overhead and achieve balanced workloads when document length disparity is moderate.

\begin{algorithm}[t]
    \small
    \caption{Heuristic sharding algorithm of {\Mname}.} \label{alg:sharding}
\begin{algorithmic}
    \STATE \textbf{Input:} inputs $\mathbf{D}=[d_1,d_2,\cdots, d_n]$, Target Imbalance Ratio $R$ \\
    \STATE \textbf{Output:} Sharding Plan $\mathit{Per\_Doc\_P}$ and $\mathit{Whole\_Doc\_P}$ \\
\end{algorithmic}
\begin{algorithmic}[1]
    \vspace{0.5em}
    \STATE Sort $\mathbf{D}$ in descending order of their lengths.
    \STATE Initialize empty per-doc sharding plan $\mathit{Per\_Doc\_P}$  
    \STATE Initialize current imbalance ratio  $\mathit{Cur\_R} = \mathit{Inf}$  
    \WHILE{$\mathit{Cur\_R}$ $>$ $R$}
        \STATE Initialize empty temporary whole-doc sharding plan $\mathit{tmp\_P}$  

        \FOR{doc $d$ in $\mathbf{D}$}
            \STATE \BlueComment{Add $d$ to the least-loaded worker:}
            \STATE $tmp\_P.\mathit{Min\_Worker\_Add}(d)$
        \ENDFOR

        \vspace{0.5em}
        \STATE \BlueComment{Make sure Eq.~\ref{eq:2} is satisfied:}
        \WHILE{$tmp\_P$ is not equal token}
            \STATE \BlueComment{Pop documents from the worker with the most tokens:}
            \STATE $docs = tmp\_P.\mathit{Max\_Token\_Worker\_Pop}()$
            \STATE \BlueComment{Apply Whole-Doc sharding:}
            \STATE $tmp\_P.\mathit{Whole\_Doc\_Shard\_and\_Add}(docs)$
        \ENDWHILE

        \vspace{0.5em}
        \STATE \BlueComment{Update current imbalance ratio:}
        \STATE $Cur\_R=T.\mathit{Compute\_Imba\_Ratio}(tmp\_p)$
        \IF{$Cur\_R$ $>$ $R$}
            \STATE \BlueComment{Pop the longest document and apply Per-Doc sharding:}
            \STATE $d = \mathbf{D}.\mathit{Pop\_Front}()$
            \STATE $Per\_Doc\_P.Add(d)$
        \ENDIF
    \ENDWHILE

    \vspace{0.5em}
    \STATE $\mathit{Whole\_Doc\_P} = tmp\_P$ 
    \STATE \textbf{Return} $\mathit{Per\_Doc\_P}$ and $\mathit{Whole\_Doc\_P}$
\end{algorithmic}
\end{algorithm}

\begin{figure*}[h]
    \captionsetup[sub]{font=small, labelfont=small}
      \centering
      \begin{subfigure}[b]{0.47\textwidth}
        \centering
        \includegraphics[width=\textwidth]{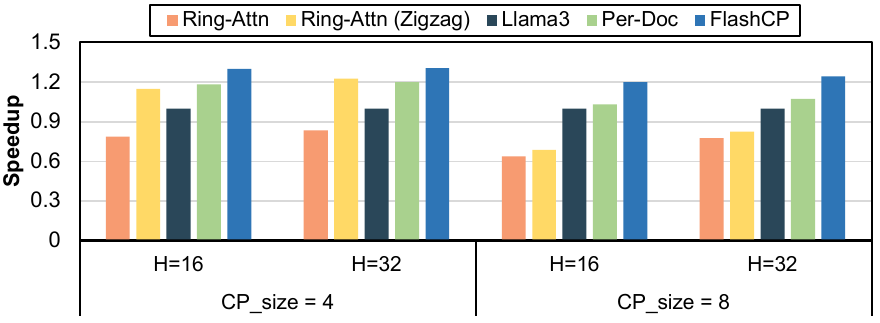}
        \vspace{-15pt}
        \subcaption{Inference latency comparison on \textit{WLB-LLM} dataset.}
      \end{subfigure}
      \begin{subfigure}[b]{0.47\textwidth}
        \centering
        \includegraphics[width=\textwidth]{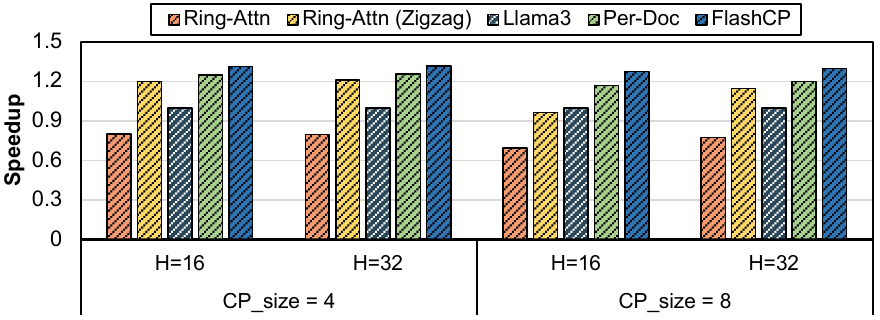}
        \vspace{-15pt}
        \subcaption{Training latency comparison on \textit{WLB-LLM} dataset.}
      \end{subfigure}

      \vspace{3pt}
    
      \begin{subfigure}[b]{0.47\textwidth}
        \centering
        \includegraphics[width=\textwidth]{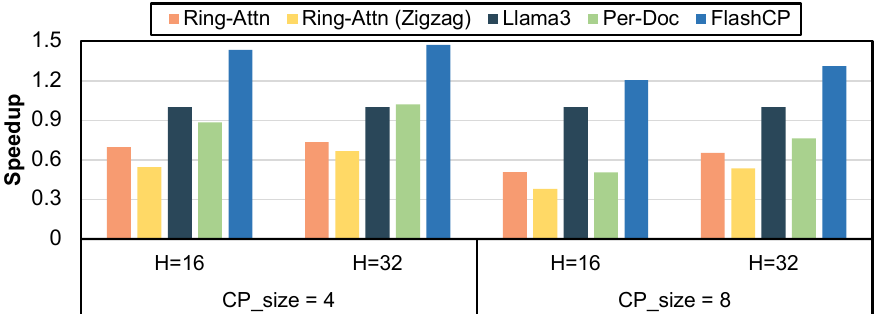}
        \vspace{-15pt}
        \subcaption{Inference latency comparison on \textit{Pile} dataset.}
      \end{subfigure}
      \begin{subfigure}[b]{0.47\textwidth}
        \centering
        \includegraphics[width=\textwidth]{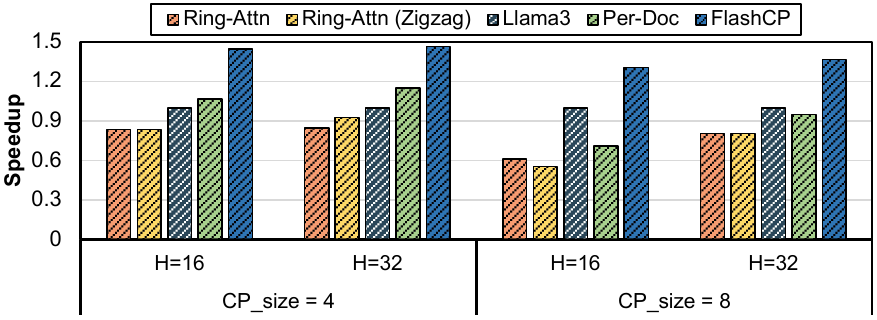}
        \vspace{-15pt}
        \subcaption{Training latency comparison on \textit{Pile} dataset.}
      \end{subfigure}

      \vspace{3pt}
    
      \begin{subfigure}[b]{0.47\textwidth}
        \centering
        \includegraphics[width=\textwidth]{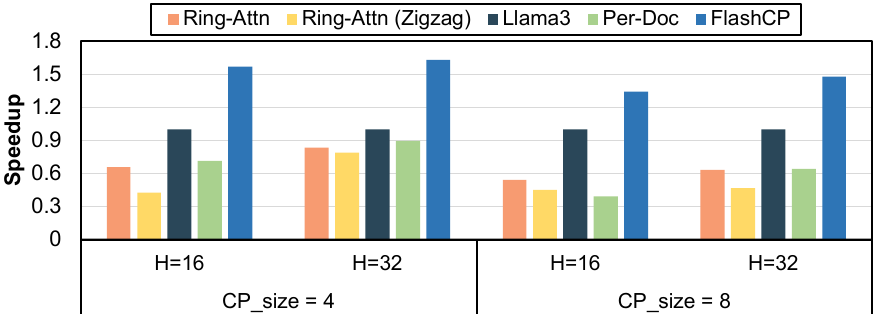}
        \vspace{-15pt}
        \subcaption{Inference latency comparison on \textit{RedPajama} dataset.}
      \end{subfigure}
      \begin{subfigure}[b]{0.47\textwidth}
        \centering
        \includegraphics[width=\textwidth]{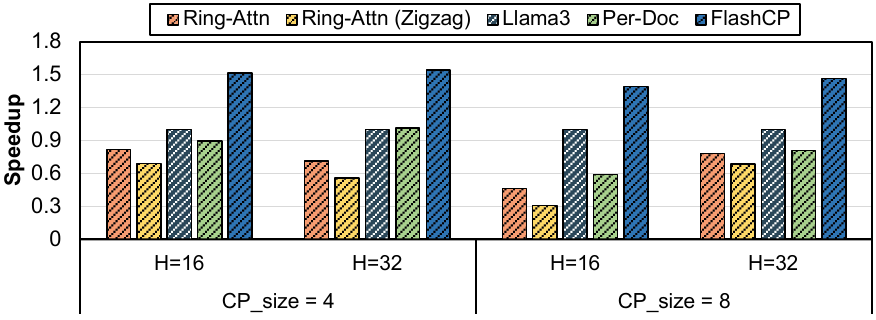}
        \vspace{-15pt}
        \subcaption{Training latency comparison on \textit{RedPajama} dataset.}
      \end{subfigure}
    
      \caption{CP training and inference performance comparison: (a)(b) are on \textit{WLB-LLM}, (c)(d) are on \textit{Pile}, and (e)(f) are on \textit{RedPajama}. $H$ refers to the number of attention heads. The context window size is set to 128K and the head dimension is 128. The speedup data are normalized to the latency of \textit{Llama3 CP} baseline.}
      \label{fig:main_result}
      \vspace{-10pt}
    \end{figure*}

\subsection{{\Mname} Sharding Algorithm}\label{sec:sharding_algo}

\textbf{ILP-based Sharding:} Based on the problem formulation in Section~\ref{sec:formulation}, we can model the sharding task as an integer linear programming (ILP) problem by incorporating the communication overhead term (Eq.~\ref{eq:5}) into the objective function. This formulation allows us to employ ILP solvers to obtain an optimal input sharding and distribution plan for a given sharding granularity. However, the computational cost of solving the ILP is prohibitively high for practical use, which motivates the design of a more efficient heuristic sharding algorithm to search for a near-optimal solution.

\textbf{{\Mname} Heuristic Sharding Algorithm:} To efficiently search for a near-optimal sharding plan, we propose a greedy heuristic sharding algorithm. The details of the algorithm are given in Algorithm~\ref{alg:sharding}.
The algorithm takes the document sequence $\mathbf{D} = [d_1, d_2, \cdots, d_n]$ and the target imbalance ratio $R = \frac{{max\_workload}}{{avg\_workload}}$ as input. Here, ${max\_workload}$ and ${avg\_workload}$ are the maximum and average attention computation workloads across all CP workers, respectively.
The algorithm first sorts documents by decreasing length, then iteratively constructs a temporary plan $tmp_p$ by assigning each document to the worker with the minimum workload (lines 5–9). During this process, the algorithm distributes entire documents to specific CP workers without sharding, which helps maximize communication savings and maintain kernel efficiency.
After that, the algorithm checks the number of tokens assigned to each worker. If the token counts are not equal, it applies \textit{Whole-Doc} sharding to balance both the token distribution and the attention computation workload (lines 10–16).
After that, $tmp\_p$ becomes a valid sharding plan that satisfies the equal-token constraint (Eq.~\ref{eq:2}), and the algorithm computes the imbalance ratio $Cur\_R$ of the current sharding plan $tmp\_p$ (line 18).
If the imbalance ratio of $tmp\_p$ is larger than $R$, it indicates that there may be some extremely long sequences that hinder load balancing. To address this issue, the algorithm removes the longest document from the input sequence and applies \textit{Per-Doc} sharding to it (lines 19–23). The remaining documents are then used in the subsequent iterations. 
This process is repeated until the achieved imbalance ratio $Cur\_R$ becomes less than the target ratio $R$. The algorithm then returns $tmp\_p$ as the final \textit{Whole-Doc} sharding plan, along with the set of documents assigned to \textit{Per-Doc} sharding.

\section{Experiments}

\subsection{Experiment Setup}\label{sec:experiments_setup}

\textbf{Experiment Environments:} All of our experiments are conducted on a single node with $8\times$ NVIDIA H100 SXM 80GB GPUs interconnected via high-bandwidth NVLink.

\textbf{Baselines:} We compare \textit{\Mname} against several state-of-the-art CP training frameworks:
\begin{itemize}
    \item \textit{Ring-Attn}~\cite{liu2023ring}: The pioneering CP method using P2P communication to exchange KV and overlap computation and communication. As the original version does not support input packing, we adopt an improved open-source version~\cite{ring2025zigzag}.
    
    \item \textit{Ring-Attn (Zigzag)}~\cite{ring2025zigzag}: An enhanced variant that uses fine-grained sharding to improve workload balance while retaining ring-based communication.
 
    \item \textit{Llama3 CP}~\cite{dubey2024llama}: The CP framework used in Llama3, which uniformly shards the whole input sequences and employs collective communication primitives (\texttt{AllGather}/\texttt{ReduceScatter}) for KV tensor and gradient synchronization.
    
    \item \textit{Per-Doc CP}~\cite{wang2025wlb}: A advanced version of \textit{Llama3 CP} that adopts per-document sharding to achieve better load balance and end-to-end efficiency, while using a similar communication pattern.
\end{itemize}

\begin{figure*}[t]
    \captionsetup[sub]{font=small, labelfont=small}
      \centering
      \begin{subfigure}[b]{0.45\textwidth}%
        \centering
        \includegraphics[width=\textwidth]{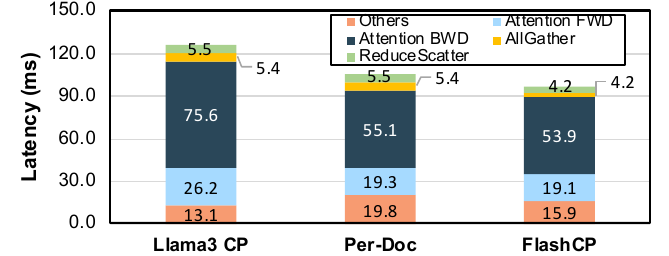}
        \subcaption{Latency breakdown on \textit{WLB-LLM} dataset.}
      \end{subfigure}%
      \hspace{5pt}
      \begin{subfigure}[b]{0.45\textwidth}%
        \centering
        \includegraphics[width=\textwidth]{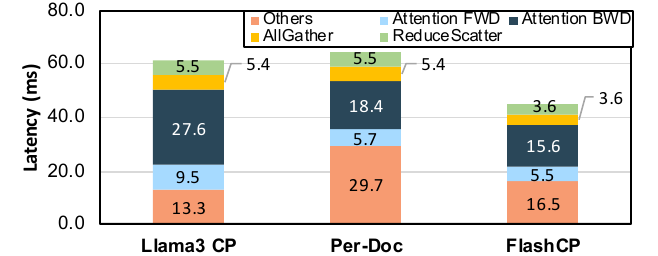}%
        \subcaption{Latency breakdown on \textit{Pile} dataset.}
      \end{subfigure}
    
      \caption{The training latency breakdown of \textit{\Mname} and two baselines on the \textit{WLB-LLM} and \textit{Pile} datasets. The data are collected from the intra-node experiments with 8 CP workers.}
      \label{fig:breakdown}
      \vspace{-10pt}
    \end{figure*}

\textbf{Datasets.} We evaluate \textit{\Mname} on three LLM datasets: 
\begin{itemize}
    \item \textit{WLB-LLM}~\cite{wang2025wlb}: We use the document length distribution released in the \textit{WLB-LLM} paper and randomly generate documents that follow this distribution. The data distribution was collected from production-level training data used by Meta.

    \item \textit{Pile}~\cite{gao2020pile}: The Pile is a large-scale English text dataset designed for training large language models. It has been widely used in many open-source LLM training~\cite{gpt-neo,touvron2023llama}.

    \item \textit{RedPajama}~\cite{together2023redpajama}: RedPajama is built by Together.ai and is an open-source recipe for reproducing the LLaMA training dataset. It consists of 1.2 trillion tokens drawn from diverse sources such as CommonCrawl, C4, GitHub, arXiv, and more.
\end{itemize}

Due to the large size of each dataset, we randomly sample 100K input sequences from each for evaluation. Each input sequence is composed of multiple documents. If the total length of the input documents exceeds the context window size, the last document is truncated to fit within the limit.

\begin{figure*}[t]
\captionsetup[sub]{font=small, labelfont=small}
  \centering
  \begin{subfigure}[b]{0.43\textwidth}%
    \centering
    \includegraphics[width=\textwidth]{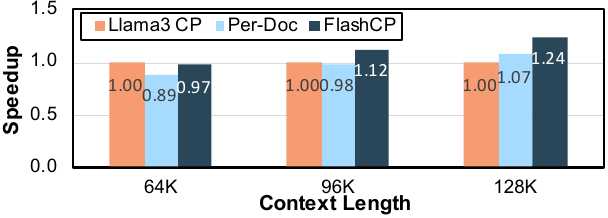}%
    \vspace{-5pt}
    \subcaption{CP inference speedup.}
  \end{subfigure}%
  \hspace{15pt}
  \begin{subfigure}[b]{0.43\textwidth}%
    \centering
    \includegraphics[width=\textwidth]{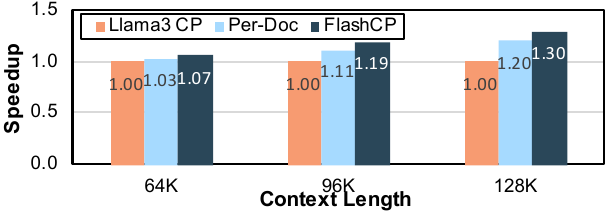}%
    \vspace{-5pt}
    \subcaption{CP training speedup.}
  \end{subfigure}

  \caption{Speedups of \textit{\Mname} across different context window sizes. The data are collected from the intra-node experiments with 8 CP workers using the \textit{WLB-LLM} dataset.}
  \label{fig:scale_context}
  \vspace{-10pt}
\end{figure*}

\subsection{Training and Inference Performance Comparison}
We evaluate the CP training and inference performance of \textit{\Mname} and all baselines on three benchmark datasets, using two model configurations with 16 and 32 attention heads (head dimension 128) and a context window of 128K. We experiment with CP sizes of 4 and 8.

\textbf{Comparison to \textit{Llama3 CP} and \textit{Per-Doc CP}:} As shown in Figure~\ref{fig:main_result}, \textit{\Mname} consistently outperforms both baselines across various configurations, achieving average speedups of $1.38\times$ over \textit{Llama3 CP} and $1.63\times$ over \textit{Per-Doc CP}.
Compared to \textit{Llama3 CP}, the performance gain mainly arises from improved workload balance. \textit{Llama3 CP} uniformly shards the input sequence, leading to load imbalance. In contrast, \textit{\Mname} adopts a hybrid, adaptive sharding strategy that distributes input documents in a workload-aware manner, achieving significantly better load balancing.
Compared to \textit{Per-Doc CP}, the improvement stems from reduced communication overhead and higher kernel efficiency. Although \textit{Per-Doc CP} achieves load balance via fine-grained sharding, it incurs extra overhead from inefficient attention kernels and frequent small data transfers.
\textit{\Mname} addresses this by applying a \textit{Whole-Doc} sharding strategy that maintains good workload balance while avoiding unnecessary sharding of short input documents. Additionally, its dynamic communication optimization further reduces communication overhead, significantly improving the overall performance.

\textbf{Comparison to \textit{Ring-Attn} and \textit{Ring-Attn (Zigzag)}:} \textit{\Mname} achieves an average speedup of $1.97\times$ over the \textit{Ring-Attn} baseline and $2.14\times$ over the \textit{Ring-Attn (Zigzag)} baseline. Although both \textit{Ring-Attn} and \textit{Ring-Attn (Zigzag)} overlap communication with computation to mitigate communication overhead, they still suffer from low kernel efficiency. This inefficiency arises from two main factors. First, the ring-based attention mechanism requires splitting the original attention kernel into multiple smaller blockwise kernels, which reduces overall kernel utilization. Second, since each kernel computes only a partial attention result, both methods require an additional partial result processing steps to obtain the final attention output.

\textbf{Across different datasets:} \textit{Per-Doc CP} and \textit{Ring-Attn (Zigzag)} perform well on \textit{WLB-LLM} but degrade significantly on \textit{Pile} and \textit{RedPajama}. The reason is the difference in document length distribution across dataset. \textit{WLB-LLM} is more skewed with extremely long documents, while \textit{Pile} and \textit{RedPajama} contain more shorter sequences. Both \textit{Per-Doc CP} and \textit{Ring-Attn (Zigzag)} employs fine-grained per-document sharding strategy, which will reduce the input shard length and decrease the kernel efficiency especially when the input sequence is mostly consists of multiple short documents. In contrast, \textit{\Mname} achieves consistently strong performance across all datasets, demonstrating the robustness of its adaptive sharding strategy.

\subsection{Optimization Analysis}

\textbf{Training Latency Breakdown:}
To analyze the performance gains of \textit{\Mname} and the impact of each individual optimization, we present a training latency breakdown of \textit{\Mname} and two baselines on \textit{WLB-LLM} and \textit{Pile}. Results are shown in Figure~\ref{fig:breakdown}.
For communication latency (AllGather and ReduceScatter), \textit{Llama3 CP} and \textit{Per-Doc CP} incur similar costs due to full KV tensor exchange. In contrast, \textit{\Mname} only communicates the required portions of the KV tensor for each CP worker, reducing latency by $23.6\%$ and $34.5\%$ on \textit{WLB-LLM} and \textit{Pile}, respectively.
Regarding attention kernel latency, both \textit{Per-Doc CP} and \textit{\Mname} outperform \textit{Llama3 CP} due to improved workload balance. Moreover, \textit{\Mname} further reduces latency over \textit{Per-Doc CP} by avoiding unnecessary fine-grained sharding, thereby preserving higher kernel efficiency.
The remaining portion of the latency (labeled as \textit{Others}) comes from data copy overhead. \textit{Per-Doc CP} introduces significantly higher data copy latency, since its fine-grained sharding produces many small KV tensor copies. Instead, \textit{\Mname} introduces only modest overhead by selectively sharding long documents while keeping short ones intact.

\textbf{Speedup Across Context Window Sizes:} We investigate the impact of context window size on the performance gains delivered by \textit{\Mname{}}. We evaluate \textit{\Mname{}} and all baselines across context window sizes ranging from $64\text{K}$ to $128\text{K}$. As shown in Figure~\ref{fig:scale_context}, \textit{\Mname{}} outperforms both baselines across almost all context window size configurations. Moreover, as the context window size increases, the speedup achieved by \textit{\Mname{}} becomes more pronounced.
This is because the attention computation cost increases quadratically with context length, causing larger context windows to exacerbate workload imbalance across CP workers. This further highlights the effectiveness of \textit{\Mname{}} in mitigating such imbalance.
This increasing trend in speedup demonstrates the strong potential of \textit{\Mname{}} in long-context LLM training, especially given the ongoing trend toward ever-larger context window sizes.

\begin{table}[t]
    \centering
    \small
    \caption{Comparison between ILP Solver and Heuristic Algorithm.}
    \vspace{5pt}
    \label{tab:ilp_flashcp}
    \begin{tabular}{c|cc}
    \toprule
    \textbf{Metric} & \textbf{ILP Solver} & \textbf{Heuristic } \\
    \midrule
    Communication Saving     & $36\%$ & $28\%$ \\
    Workload Imbalance Ratio & 1.00 & 1.04 \\
    \bottomrule
    \end{tabular}
    \vspace{-10pt}
\end{table}

\textbf{Optimality Study:} We evaluate the optimality of our heuristic sharding algorithm by comparing its communication saving and workload imbalance ratios against an ILP solver on the \textit{Pile} dataset with 4 CP workers. The workload imbalance ratio is defined as $\frac{\textit{max\_workload}}{\textit{avg\_workload}}$, where $\textit{max\_workload}$ and $\textit{avg\_workload}$ denote the maximum and average attention workloads across all CP workers, respectively. As shown in Table~\ref{tab:ilp_flashcp}, the ILP solver achieves a communication saving ratio of $36\%$ and a workload imbalance ratio of $1.00$. However, it incurs high computational cost. Solving a single input sequence can take tens of minutes, making it impractical for real-world use. 
In contrast, our heuristic achieves $28\%$ communication saving and a $1.04$ imbalance ratio, both very close to those of the ILP solver. These results demonstrate that the heuristic algorithm effectively balances communication savings and workload distribution, achieving near-optimal performance.

\section{Related Works}

As LLM context windows continue to grow~\cite{team2023gemini, team2024gemini, singh2025meta, zhu2025skyladder}, training becomes increasingly constrained by the rapidly growing activation memory of attention layers~\cite{korthikanti2023reducing, wang2025lemo}. CP has recently emerged as an effective strategy that partitions input sequences and activations along the sequence dimension, distributing attention computation across GPUs~\cite{nvidia2023megatron, gu2024loongtrain, dubey2024llama, wang2025wlb}.
Early CP methods adopt ring-based communication to exchange KV tensors among workers~\cite{liu2023ring}, enabling partial overlap of computation and communication but struggling to support the complex attention masks required by input packing. To address this limitation, zigzag-style sharding is introduced~\cite{ring2025zigzag}. However, it still suffers from reduced kernel efficiency due to blockwise attention execution.
More recent approaches leverage collective communication primitives, such as AllGather~\cite{nvidia2023megatron, dubey2024llama} and AlltoAll~\cite{jacobs2023deepspeed}, to enable efficient attention computation with document masking by providing each worker with a global KV view. However, these methods communications the entire KV tensor, introducing significant redundant communication.

\section{Conclusion}\label{sec:conclusion}

In this paper, we introduce \textit{\Mname}, a load-balanced and communication-efficient context parallelism framework for large-scale, long-context LLM training. Specifically, \textit{\Mname} proposes a sharding-aware communication mechanism that minimizes unnecessary communication, and a novel \textit{Whole-Doc} sharding strategy to maximize communication savings while maintaining balanced workload distribution across CP workers. Moreover, \textit{\Mname} further introduces a heuristic sharding algorithm to efficiently search for a near-optimal sharding plan. Extensive experiments show that \textit{\Mname} delivers up to $1.63\times$ speedup over state-of-the-art CP frameworks across diverse datasets.

\bibliography{ref}
\bibliographystyle{icml2026}




\end{document}